\documentclass{ws-ijmpb}
\usepackage[latin1]{inputenc}
\usepackage[T1]{fontenc}
\usepackage{bm}
\usepackage{graphicx}
\usepackage{amsmath}
\usepackage{amsfonts}
\usepackage{amssymb}
\usepackage{color}
\usepackage{multirow}
\usepackage[super,comma,sort]{natbib}
\setcitestyle{super}

\usepackage{epstopdf}

\newcommand*{\onlinecite}[1]{%
  \begingroup
    \romannumeral-`\x 
    \setcitestyle{numbers}%
    \cite{#1}%
  \endgroup   
}

\begin{document}
\title{TUNABLE SPIN DEPENDENT TRANSPORT IN THE DOT--RING NANOSTRUCTURE}
\author{BARBARA K\k{E}DZIERSKA, ANNA GORCZYCA--GORAJ, EL\.ZBIETA ZIPPER, MACIEJ~M.~MA\'SKA}
\address{Institute of Physics, University of Silesia, 75 Pu{\l}ku Piechoty 1, 41-500 Chorz\'ow, Poland}

\maketitle

\begin{abstract}
We study spin--dependent transport through a quantum nanostructure composed of a quantum dot surrounded by a quantum nanoring. The nanostructure is side--attached to source and drain electrodes and we assume that the confining potential that forms the nanostructure can be controlled by electrical gating. We demonstrate that depending on the model parameters the system can exhibit positive as well as negative differential conductance. What is more interesting, these properties can be easily controlled by tuning the applied gate voltage.
\end{abstract}


\section{Introduction}\label{sec1}

The ability to control the quantum state of a single electron is at the heart of many developments, e.g., in
spintronics and quantum computing. Motivated by the great interest in nanospintronics we briefly discuss the
effect of spin on the transport properties in complex nanostructure. The study of individual
spins is caused by possible applications in miniaturized spintronics named single spintronics.\cite{hans,elzerman}
We consider below a quantum nanostructure composed of a semiconductor quantum dot (QD) surrounded by a quantum
ring (QR) (called hereafter Dot-Ring Nanostructure, DRN) in the Coulomb blockade regime. Such structure has been recently fabricated
by pulse droplet epitaxy.\cite{somaschini,somaschini2} 
It can be also grown employing core-shell nanowire growth\cite{lauhon, dillen} of, e.g., (In,Ga)As where the
core and shell parts are separated by the tunneling barrier. Then by cutting a slice of it one can form
a DRN. 
In general, modern sophisticated technics allow one to make DRN's in which the inner QD and the outer QR parts
are made of different materials.\cite{mohseni,sadowski} The properties of DRN can be strongly modified by changing
the shape and the height of the barrier separating the QD and QR parts and/or the relative position of
the minima of the potential wells.\cite{zipper, MK} It was shown\cite{zipper} that by changing the
confinement potential, e.g., by the electrical gating one can change the shape and distribution of the
electron wave functions which determine many physical, measurable quantities such as relaxation times,
optical and transport properties. 

In this paper, we want to incorporate spin to the description and study the influence of spin on the
transport properties of DRN. Owing to the strong confinement of electrons in DRNs the orbital states
are strongly quantized which leads to a drastic reduction of spin-phonon interaction mediated by a
combination of electron-phonon and spin-orbit coupling. Thus, the electron spin states are very stable
due to the substantial suppression of spin-flip mechanisms\cite{elzerman} and the spin relaxation rates
are much smaller than the orbital relaxation rates. 

Transport through the nanostructure depends crucially on the tunneling rates of its states to the leads.
In the case of DRN, this coupling depends on the localization of the electron wave function: states
localized in QD (QR) are weakly (strongly) coupled to the electrodes.\cite{mk} To obtain spin polarized
current we apply the magnetic field $B$ parallel to the DRN plane and we design DRN in which there exist
spin dependent tunnel rates. 
In particular, we formulate conditions under which DRN exhibits 
spin related positive differential conductance (PDC) or negative differential conductance (NDC),\cite{britnell,bulka} a phenomenon widely exploited in electronic devices, for instance in high--frequency oscillators, multipliers, analog-to-digital converters etc. \cite{chen}
These features are mostly determined by the tunneling rates,\cite{hans} which reflect the shape and distribution of the wave functions and can be largely modified by the electrical gating. Thus the microscopic properties of a DRN can be engineered on demand, depending on a particular application of the DRN.

A spin blockade in single electron transistor in QD resulting from spin polarized leads has been discussed in Ref. \onlinecite{hawrylak}. 
In this paper we outline another mechanism for spin dependent current which utilizes the peculiar properties of DRN in the presence of both spin unpolarised and polarised leads. We present below the results for the simple case of DRN occupied by a \textit{single} electron.

\section{Dot--Ring Nanostructure}\label{sec2}

We consider a 2D, circularly symmetric dot-ring nanostructure defined by a confinement potential $V(r)$. It is composed of a QD surrounded by a QR and separated from the ring by a potential barrier $V_0$ which enables the electron tunneling between the QD and QR.
\begin{figure}[h]
\begin{center}
\includegraphics[width=0.6\textwidth]{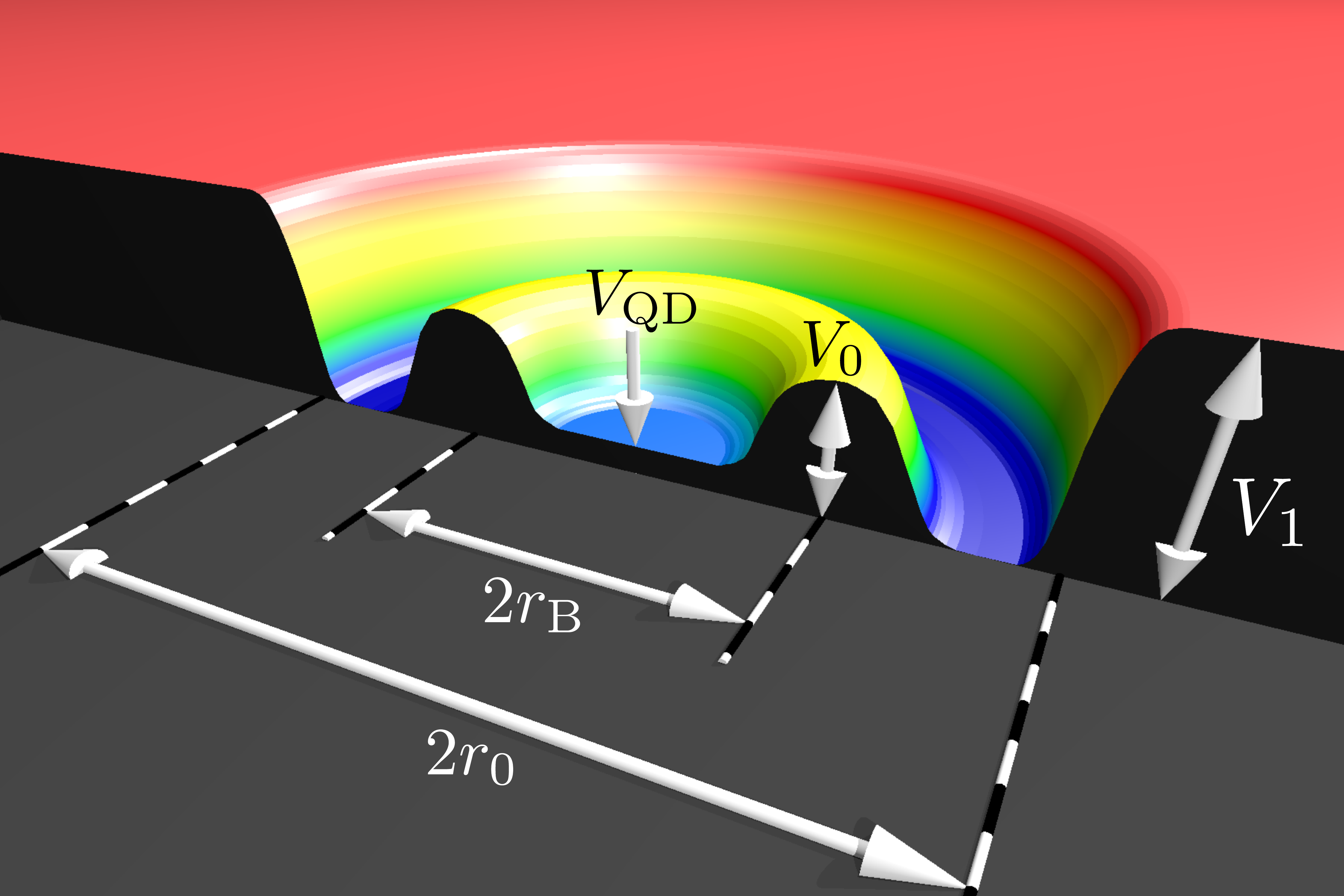}
\caption{Schematic illustration of DRN.}
\label{DRN}
\end{center}
\end{figure}
 A cross section of a DRN with explanation of symbols used throughout the text is presented in Fig.~\ref{DRN}.

The single--electron Hamiltonian in the presence of the in-plane magnetic field $B$ is written as
\begin{equation}
H = \frac{1}{2m^{*}} {\bf p}^2 + \frac{e\hbar}{2m^*} \mathbf{\hat{\sigma}} \cdot \mathbf{B}+V(r),
\label{Hamiltonian_pr}
\end{equation}

The confinement potential $V(r)$ can be controlled by electrical gating which changes $V_{\rm QD}$ - the parameter
describing the bottom of the inner QD potential. The energy spectrum of $H$, calculated by numerical solving the
Schr\"odinger equation, consists of a set of discrete states $E_{nl}$ due to radial motion with radial
quantum numbers $n=0,1,2,\ldots$, and rotational motion with angular momentum quantum numbers
$l=0,\pm 1,\pm 2\ldots$. To observe spin effects we assume that a large in-plane magnetic field $B$
is applied to DRN which has a negligible effect on orbitals, but causes substantial Zeeman splitting
$\Delta _Z=g\mu_B B$ where $g$ is the spin gyromagnetic factor.
The single particle wave function is of the form 
\begin{equation}
\Psi_{nl\sigma} = R_{nl}\left(r\right)\exp\left(i l \phi \right)\chi_{\sigma},
\label{eq_psi_nl}
\end{equation}
with the radial part $R_{nl}(r)$ and the spin part $\chi_{\sigma}$.
In our model calculations we take the radius of DRN $r_0=50$ nm, $V_1=90$ meV and set the zero potential energy at the level of $V_{\rm QR}$, i.e., the potential well offset is equal $V_{\rm QD}$. We consider here a case of a single electron trapped in DRN (although we have also discussed the  many-particle states for $N_e=2$ i $N_e=3$\cite{biborski}). The two lowest energy levels are ($E_{0} \equiv E_{00}$):

\begin{equation}
E_0^{\uparrow\downarrow} = E_0 \pm \frac{g\mu_{B}B}{2}.
\label{spin}
\end{equation} 

\begin{figure}[h]
\begin{center}
\includegraphics[width=0.7\textwidth]{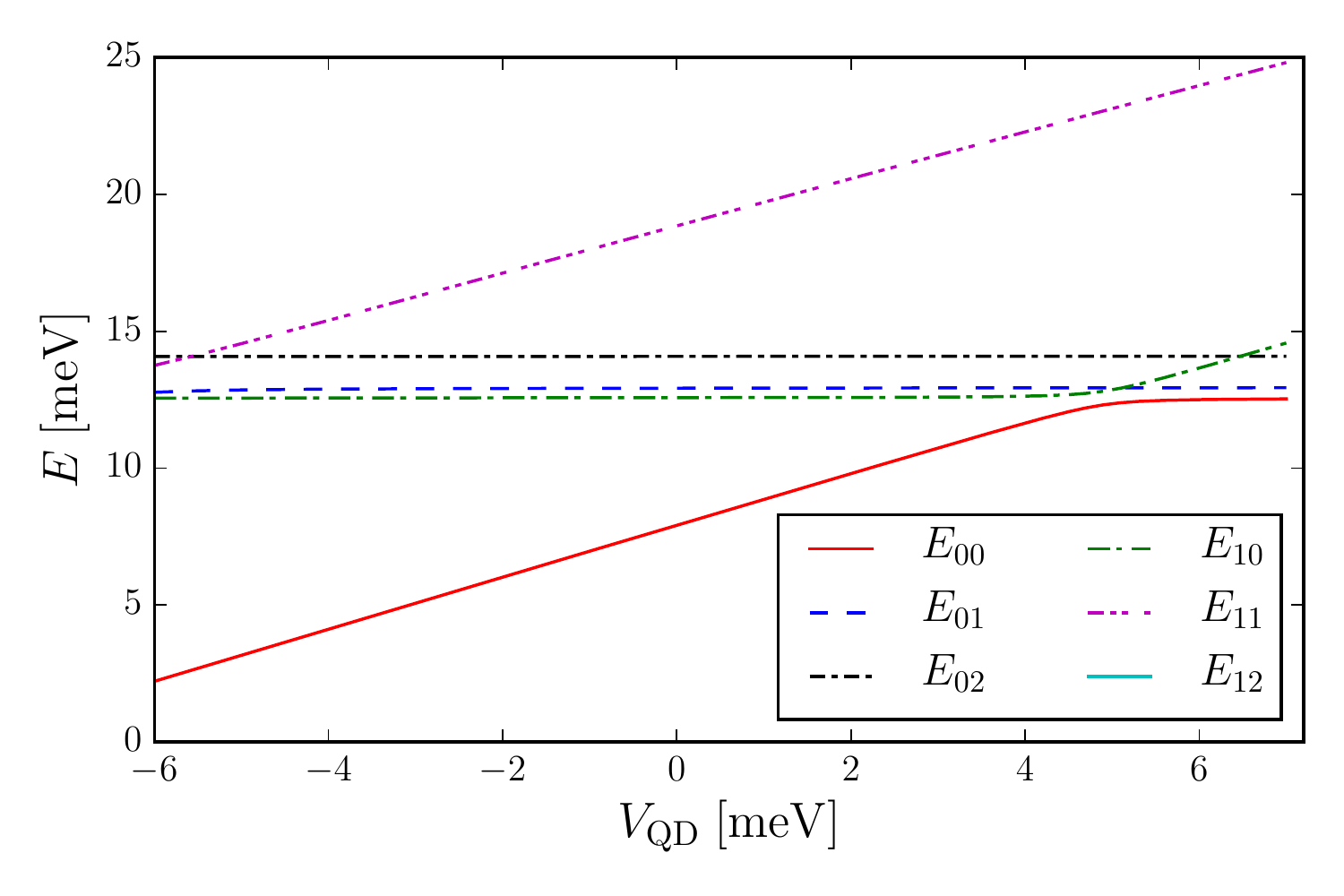}
\end{center}
\caption{The energy spectrum of DRN. \label{spectrum}}
\end{figure}

The energy spectrum as a function of $V_{\rm QD}$ in the absence of magnetic field is shown in Fig. \ref{spectrum}.
Depending on the shape of the confining potential, controlled by the value of $V_{\rm QD}$, particular states are
located in QD or QR part of the DRN. If a state is in the QD, its energy almost linearly increases with increasing 
$V_{\rm QD}$, whereas the energies of states in the QR are not affected. In this manner even without the knowledge 
of the precise shape of the wave function one is able to determine if the electron is located in the QD or the QR.

\section{Negative and Positive Differential Conductance}

The system we study is a DRN weakly coupled via tunneling barriers to the source and drain electrodes with
chemical potentials $\mu_{\rm S}$ and $\mu_{\rm D}$, respectively. We estimate the sequential tunneling current $I$
from the source electrode through DRN to the drain electrode in the Coulomb blockade (CB) regime, i.e., the
current is restricted to flow one electron at a time. To calculate the current we make use of the steady state
solutions of rate equations\cite{mk}
for the average electron number $\rho_i$ at the $i$-th energy state which is situated in the bias window
$\mu_{\rm S}-\mu_{\rm D} = eV_{\rm SD}$, where $V_{\rm SD}$ is the bias voltage,
and eventually obtain

\begin{equation}
  \left(f_{\rm S}^{i}\Gamma_{\rm S}^{i}+f_{\rm D}^{i}\Gamma_{\rm D}^{i}\right)\left(1-\sum_{j}\rho_j\right)-\left[(1-f_{\rm S}^{i})\Gamma_{\rm S}^{i}
    +(1-f_{\rm D}^{i})\Gamma_{\rm D}^{i}\right]\rho_i+W_i=0
\label{rate_eq}
\end{equation}
Here, $i$ denotes all the quantum numbers of a particular state $i=nl\sigma$, ($\sigma=\uparrow \downarrow$),
$f_{\rm S,D}^{i}$ are the Fermi--Dirac distribution functions for the electrodes ($f^i_{\rm S,D}\equiv f_{\rm S,D}(E_i)$), $\Gamma_{\rm S}$, $\Gamma_{\rm D}$ are the tunnel rates
from the source electrode and to the drain electrode, respectively (calculated by Bardeen's method \cite{Bardeen}).
%
$W$ describes the change of the occupation of state $i$ due to relaxation processes. Namely, the occupation is
increased by the relaxation from states higher in energy and decreased by relaxation to states with lower energy
\begin{equation}
    W_i=\sum_{j=i+1}^{n_0-1} w_{j\rightarrow i} \rho_j\delta_{\sigma_i\sigma_j} -\sum_{j=0}^{j-1} w_{i\rightarrow j} \rho_i\delta_{\sigma_i\sigma_j},
\end{equation}
where $w_{i\to j}$ is the relaxation rate from state $i$ to state $j$, $n_0$ is the total number of states and $\sigma_i$ is the spin of an electron in state $i$. The presence of $\delta$'s in the above equation means that we neglect 
relaxation processes between states with different spin orientation. As a result $W_i$ includes only the
orbital relaxation, what will be explained below. 



The DRN quantum states are assigned spin-down and spin-up. The transitions between these states, due to the coupling to the environment, strongly reflect their features. Namely, the relaxation rates of an excited state in nanostructures are very different if the transition involves ($w_{i\to j}^{\rm spin}$) or does not involve spin flip ($w_{i\to j}^{\rm orb}$).\cite{bockelman,nazarov} It was found both experimentally and theoretically that the relaxation rate $w_{i\to j}^{\rm spin}$ between the energy states is much smaller than the orbital relaxation rate $w_{i\to j}^{\rm orb}$ due to electron--phonon interaction ($w_{i\to j}^{\rm orb} \gtrsim 1\rm{GHz}$).\cite{mk} Small $w_{i\to j}^{\rm spin}$ ($w_{i\to j}^{\rm spin} <10^5\rm{Hz}$) is responsible for the observation of long lived transient currents in nanostructures.\cite{Fujisawa} $w_{i\to j}^{\rm spin}$ calculated in Ref. \onlinecite{zipper} ranges from 10 Hz to $10^4$ Hz. Therefore, in the rate equations (\ref{rate_eq}) we neglect the spin relaxation processes during the transport through the DRN.

To investigate spin related transport via DRN we design the system so that it can exhibit 
 spin related negative differential conductance $G$ (NDC), i.e., the collapse of spin polarized current with an increase of the bias voltage. The reason of it is that the addition of a new conducting channel with smaller transmission probability effectively blocks the current and consequently $G= {\rm d}I/{\rm d}V_{\rm SD}<0$. 
 On the other hand, the addition of a new channel with larger transmission probability increases the current and one gets the  increase of the current with an increase in the number of available channels, namely positive differential conductance (PDC) with $G= {\rm d}I/{\rm d}V_{\rm SD}>0$. NDC region has attracted considerable attention due to its potential application in the realization of low power memory devices and logic circuits. NDC behavior is also known to offer great potential for high frequency applications as Bloch oscillators, frequency multipliers, and fast switching devices. \cite{chen,britnell,wu}

 In this section we assume that QD and QR are made of different materials with the respective gyromagnetic factors $g_{\rm QD}$, $g_{\rm QR}$. 
We solve the Schr\"odinger equation in external magnetic field for a DRN with a position-dependent $g$-factor. It is easy to estimate the energy of a particular state without any additional calculations, provided we know the
energy spectrum of a homogeneous system presented in Fig.~\ref{spectrum}. When magnetic field is applied to a
homogeneous system, its energies are shifted according to Eq. (\ref{spin}). However, in a DRN with different
$g_{\rm QD}$ and $g_{\rm QR}$ factors, the Zeeman splitting is different in the QD and QR parts. It means that effectively 
an electron in the QD (QR) will ``feel'' an effective confining potentials 
\begin{eqnarray}
V_{\rm QD}^{\rm eff}(B,\sigma)&=&V_{\rm QD}+\sigma\frac{g_{\rm QD}\mu_{\rm B}B}{2},\\
V_{\rm QR}^{\rm eff}(B,\sigma)&=&V_{\rm QR}+\sigma\frac{g_{\rm QR}\mu_{\rm B}B}{2}.
\end{eqnarray}
Then, the difference between the bottoms of the QD and QR potentials is $V_{\rm QD}\pm\frac{1}{2}\left(g_{\rm QD}-g_{\rm QR}\right)\mu_{\rm B}B$, 
where the sign ``$+$'' (``$-$'') corresponds to spin-up (spin-down) electrons.
Since in the original problem we solve the Schr\"odinger equation for the confining potential with
$V_{\rm QR}\equiv 0$,
we have to shift the energies by $\pm\frac{1}{2}g_{\rm QR}\mu_{\rm B}B$. 
Then, the spin-dependent energy spectrum for $g_{\rm QD} \ne g_{\rm QR}$ in external magnetic field can be
approximately expressed by the spectrum for $B=0$ as follows
\begin{equation}
E_n^{\sigma}(B\ne 0,\ V_{\rm QD}) = E_n\left[B=0,\ V_{\rm QD} +\sigma \frac{1}{2}\left(\left|g_{\rm QD}\right|-\left|g_{\rm QR}\right|\right)\mu_{\rm B}B\right] + \sigma \frac{1}{2}\left|g_{\rm QR}\right|\mu_{\rm B}B.
\end{equation}
This formula means that in order to obtain the dependence of a particular energy of a spin-up (spin-down) electron
on $V_{\rm QD}$ one has to shift the corresponding line in Fig. \ref{spectrum} up (down) by
$\frac{1}{2}\left|g_{\rm QR}\right|\mu_{\rm B}B$ and to the left (right) by
$\frac{1}{2}\left(\left|g_{\rm QD}\right|-\left|g_{\rm QR}\right|\right)\mu_{\rm B}B$.
This reasoning is approximate and valid only for $\left(\left|g_{\rm QD}\right|-\left|g_{\rm QR}\right|\right)\mu_{\rm B}B \ll V_0$,
but at least qualitatively illustrates the effect of magnetic field.

 We formulate below the conditions under which the arrangement of the spin energy levels is such that one gets spin related  NDC or PDC.
 We distinguish two different situations of low bias regime where only the ground state lies in the bias window $ \mu_{\rm S}-\mu_{\rm D}$ and high bias regime where also some excited states are in the bias. 
To show the idea we present below some analytical estimations of the ratio of the currents at high bias ($I_{H}$) and low bias ($I_{L}$) regimes,

 \begin{equation}
\eta\equiv\frac{I_{H}}{I_{L}},
\end{equation}
and find the conditions under which $\eta>1$ (PDC) and $\eta<1$ (NDC). We also study the influence of unequal tunneling rates $\Gamma_{\rm S}$, $\Gamma_{\rm D}$ on the discussed phenomenon (e.g. thicker barrier to the source electrode than to the drain electrode results in $\Gamma_{\rm S} < \Gamma_{\rm D}$). For simplicity in these analytical calculations we solve the rate equations for $T=0$ and consider only the two lowest energy levels. The currents calculated from the rate equations are:
 \begin{equation}
I_{H}=-e\cfrac{\Gamma_{\rm D}^{G}\Gamma_{\rm S}^{E}\Gamma_{\rm D}^{E}+\Gamma_{\rm S}^{G}\Gamma_{\rm D}^{E}\Gamma_{\rm D}^{G}}{(\Gamma_{\rm D}^{E}+\Gamma_{\rm S}^{E})\Gamma_{\rm D}^{G}+\Gamma_{\rm D}^{E}\Gamma_{\rm S}^{G}},
\end{equation}

 \begin{equation}
I_{L}=e\cfrac{\Gamma_{\rm D}^{G}\Gamma_{\rm S}^{G}}{\Gamma_{\rm D}^{G}+\Gamma_{\rm S}^{G}},
\end{equation}
where $G,E$ denote the ground and first excited energy states, respectively ($E_0^{\uparrow}$ , $E_0^{\downarrow}$ or vice versa, see Figs.~\ref{lvl1} and ~\ref{lvl2}).
To proceed with the analysis of the results we introduce the notation:
\begin{equation*}
k\equiv\cfrac{\Gamma_{\rm D}^{G}}{\Gamma_{\rm S}^{G}},\hspace*{5mm}
n\equiv\cfrac{\Gamma_{\rm S}^{E}}{\Gamma_{\rm S}^{G}} = \frac{\Gamma_{\rm D}^{E}}{\Gamma_{\rm D}^{G}}.
\end{equation*}
Parameter $k$ describes the ratio of the tunneling rates related to the thickness of the barriers to the drain and the source electrodes, whereas parameter $n$ reflects the mutual relation between the coupling strength of the higher and lower states to the electrodes. With this notation one finds

 \begin{equation}
\eta=(n+1)\frac{k+1}{k+2}.
\label{enka}
\end{equation}

 Let us discuss Eq. (\ref{enka}) in some limiting cases and look for the parameter regimes in which one
 finds PDC or NDC. This distinction is based on the relation between the values of the current in low and high bias regime described by the $\eta$ parameter. 

It follows from these considerations that NDC can not be obtained if the electron enters DRN via thicker barrier and leaves DRN via the thinner one. In the other two cases one can find the parameter regime in which NDC occurs. We see that the decrease of the current is larger in the case of asymmetric barriers ($\eta \to \frac{1}{2}$) than for the symmetric barriers ($\eta \to \frac{2}{3}$). The reason is that in the former case the thicker barrier to the 
drain electrode additionally blocks the transmission of the current. The relation between PDC and NDC regimes and the parameters $k$ and $n$ is summarized in Table \ref{tab:table1}.

\begin{table}[h!]
\tbl{Transport regimes defined by different values of the parameters $k$ and $n$.}
    {\begin{tabular}{|c|c|c|c|c|c|}
    \Hline
    $k$ &
    \multicolumn{2}{|c|}{\rule{0pt}{3ex}$\ll 1$} & 
    \multicolumn{2}{c|}{$1$} &
    \multirow{2}{*}{$\gg 1$} \\[2ex]
    \cline{1-5}
    $n$ &
    $\ll 1$ & $\gg 1$ & $\ll 1$ & $\gg 1$ & \\
    \hline\rule{0pt}{4ex}
    $\eta$ &
    $\approx\displaystyle\frac{1}{2}$ (NDC) & $\gg 1$ (PDC) & 
    $\approx\displaystyle\frac{2}{3}$ (NDC) & $\gg 1$ (PDC)& $> 1$ (PDC)\\[2ex]
    \Hline
    \end{tabular}}
    \label{tab:table1}
\end{table}

 We now illustrate the discussed phenomenon on two specific examples with two different sets of parameters. We present the results of the numerical calculations which take into account four lowest energy levels and all the subsequent orbital relaxation rates.   

\begin{figure}[h]
\begin{center}
\includegraphics[width=0.7\textwidth]{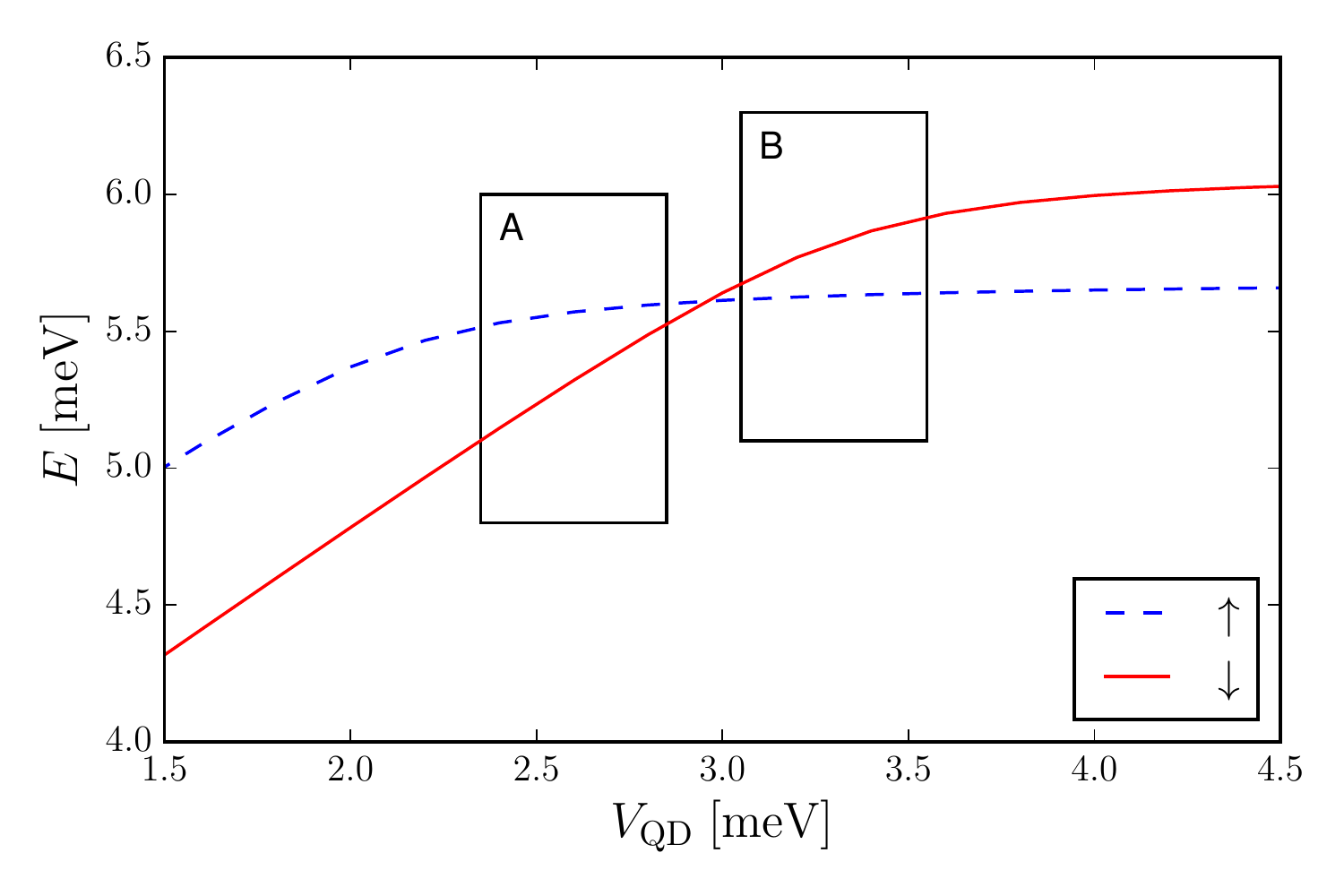}
\caption{Spin-up and spin-down energy levels as a function of $V_{\rm QD}$. Rectangles marked as A and B
show regimes, where PDC and NDC occur, respectively.}
\label{2m2_spc}
\end{center}
\end{figure}

\begin{figure}[h]
\begin{center}
\includegraphics[width=0.95\textwidth]{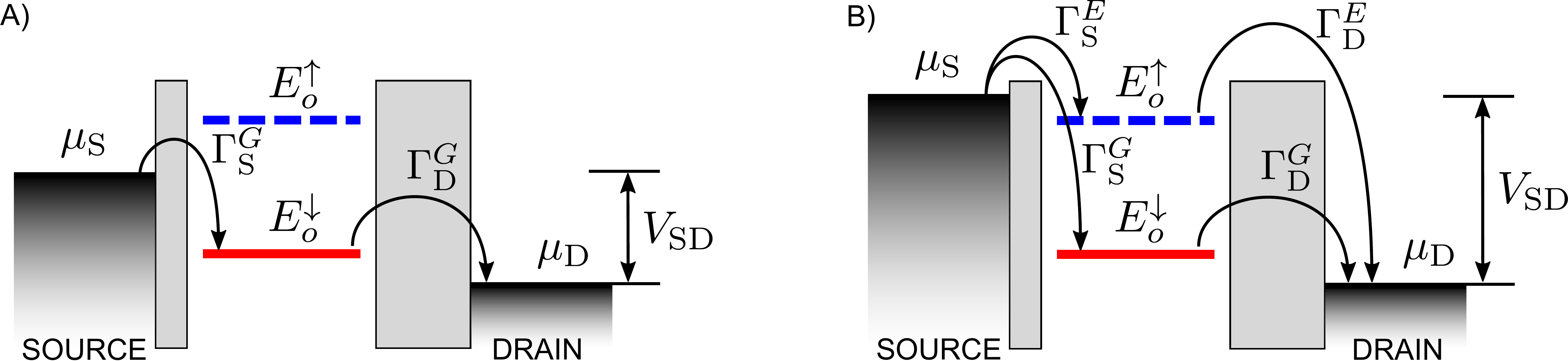}
\caption{Schematic illustration of the setup corresponding to rectangle A in Fig.~\ref{2m2_spc} for low bias (A) and high bias (B) regime. At low bias regime there is only a single weakly coupled state. Increasing potential on source opens another, strongly coupled channel.}
\label{lvl1}
\end{center}
\end{figure}

\begin{figure}[h]
\begin{center}
\includegraphics[width=0.95\textwidth]{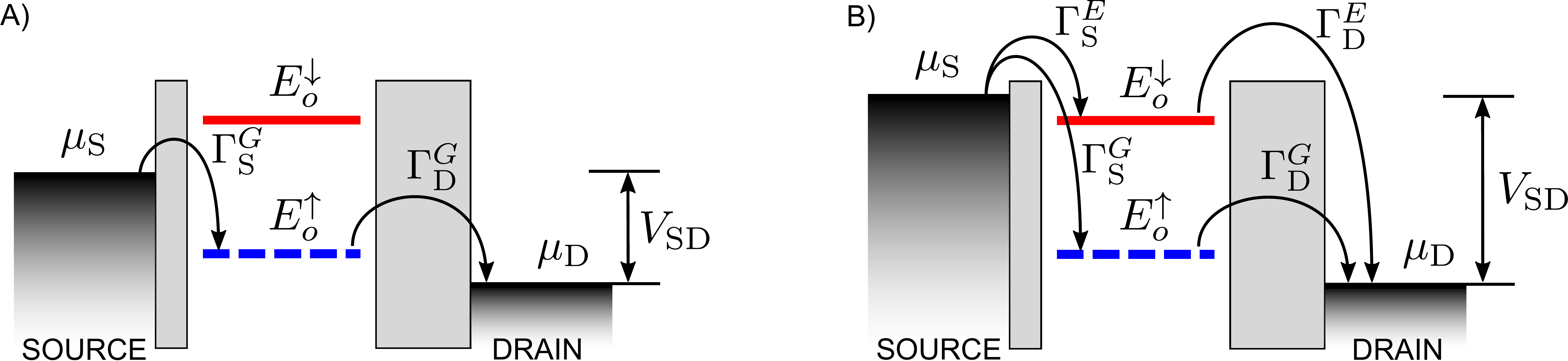}
\caption{The same as in Fig. \ref{lvl1}, but in the regime marked by rectangle B in Fig.~\ref{2m2_spc}.}
\label{lvl2}
\end{center}
\end{figure}

 At first we consider DRN in which the inner QD has gyromagnetic factor $g_{\rm QD} = -2$ and the outer QR has $g_{\rm QR}= 2$. In numerical calculations we assume $B=5$\:T. The part of the energy spectrum that is crucial for switching between PDC and NDC is shown in Fig.~\ref{2m2_spc}. 
%
%
%
Spatial location of these states can be identified by their dependence on $V_{\rm QD}$: as it was already mentioned, states situated in QD exhibit an increase in the energy with increasing $V_{\rm QD}$, whereas those situated in QR have the energy (nearly) constant.

\begin{figure}[h]
\begin{center}
\includegraphics[width=0.7\textwidth]{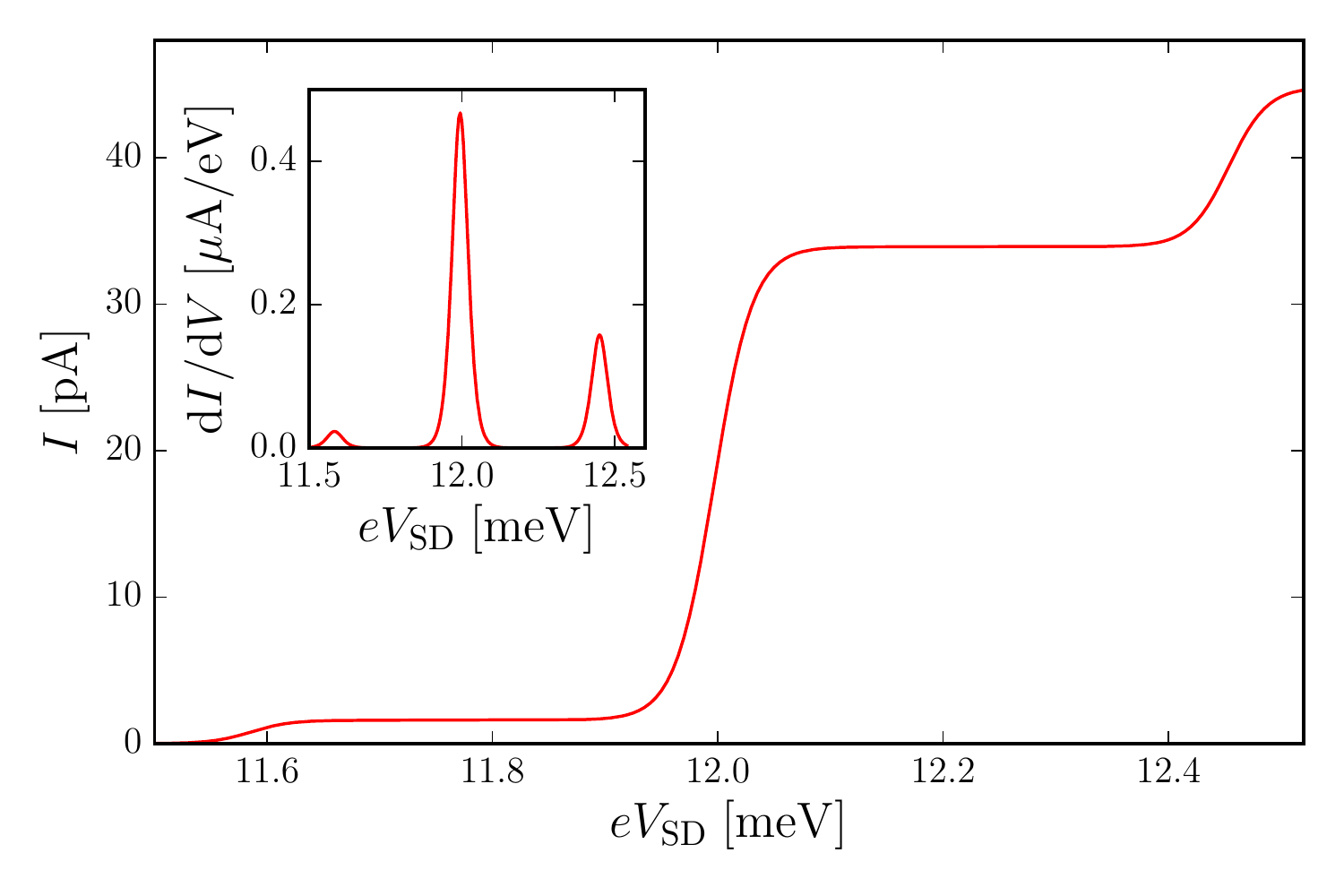}
\caption{Current as a function of applied source--drain voltage. While increasing bias window, new states are taken into account.}
\label{pdc1}
\end{center}
\end{figure}

We can distinguish between different regimes signed in Fig.~\ref{2m2_spc} with letters A and B in which we obtain spin related PDC and NDC, respectively.
In panel A (the corresponding scheme is presented in Fig.~\ref{lvl1}A) the lowest energy state $E_0^{\downarrow}$ lies in the QD with very small tunnel coupling $\Gamma_{\downarrow}$ and the energy state $E_0^{\uparrow}$ is situated mostly in the QR with much larger $\Gamma_{\uparrow}$. 
Therefore at low bias (only $E_0^{\downarrow}$ in bias window) we get weak, fully spin-down polarised current. 
At high bias ($E_0^{\downarrow}$ and $E_0^{\uparrow}$ in the bias window, see Fig.~\ref{lvl1}B) an electron can tunnel also through a spin up channel which results in an increase in the total current with a small admixture of spin down current. We will refer to this current as partially spin polarized (Fig.~\ref{pdc1}). Consequently, in this case PDC takes place.

\begin{figure}[h]
\begin{center}
\includegraphics[width=0.7\textwidth]{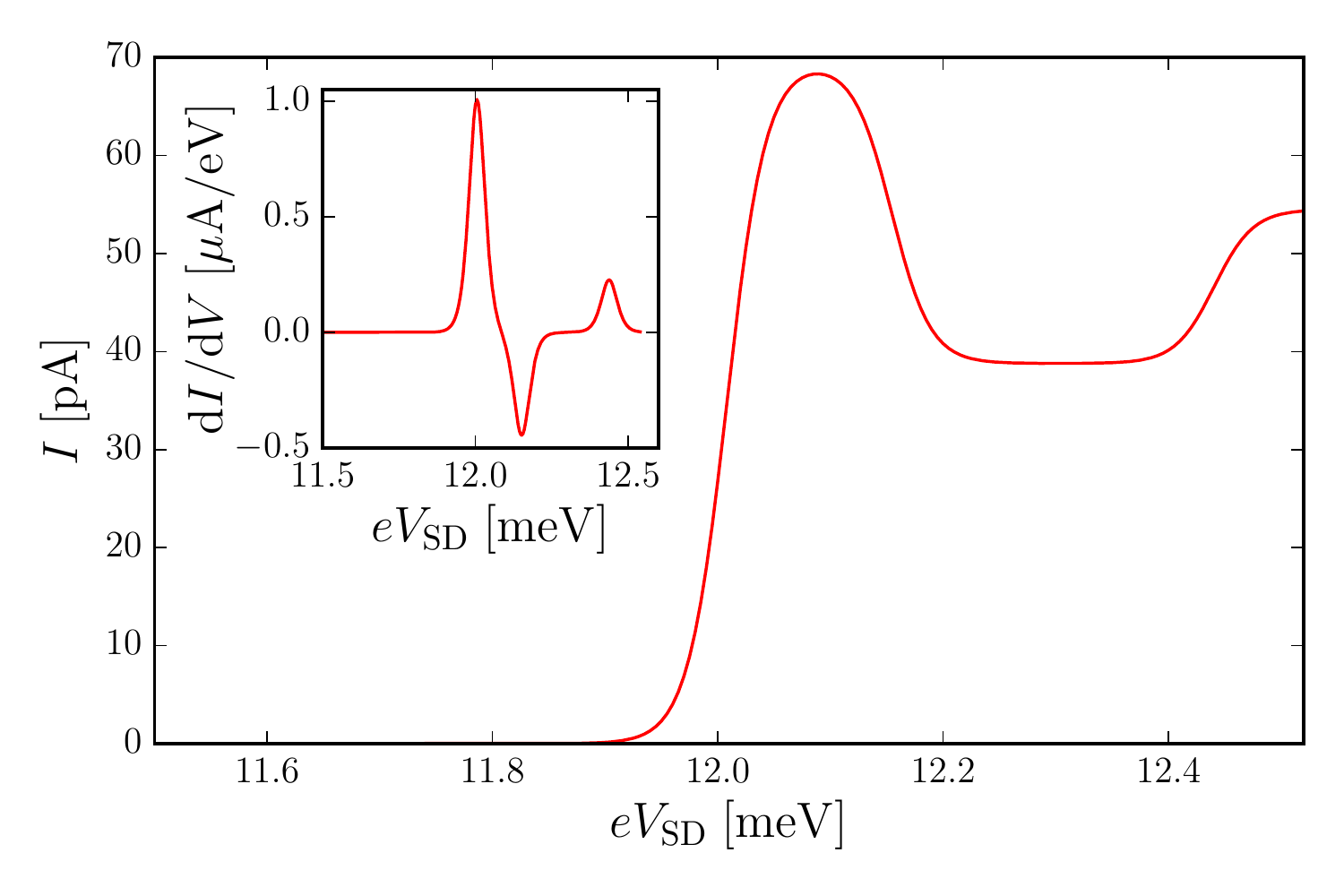}
\caption{Current as a function of applied source--drain voltage. Yet while increasing bias window, in contradistinction to Fig. \ref{pdc1}, decrease of current may occur resulting in NDC.}
\label{ndc1}
\end{center}
\end{figure}

For larger $V_{\rm QD}$, in regime marked by rectangle B in Fig. \ref{2m2_spc} (compare Fig.~\ref{lvl2}), the situation is different. The lowest energy state $E_0^{\uparrow}$ lies in the QR with large tunnel coupling
$\Gamma_{\uparrow}$ whereas the $E_0^{\downarrow}$ state is situated mostly in QD with small $\Gamma_{\downarrow}$. 
Therefore at low bias we get large, fully spin-up polarised current. At high bias ($E_0^{\uparrow}$ and $E_0^{\downarrow}$ in the bias window) an electron entering the $E_0^{\downarrow}$ state effectively blocks the spin-up channel resulting in the decrease in the (partially spin polarised) current (Fig.~\ref{ndc1}), i.e., NDC occurs.
%
By changing $V_{\rm QD}$ one can switch DRN between PDC and NDC regimes.

\begin{figure}[h]
\begin{center}
\includegraphics[width=0.7\textwidth]{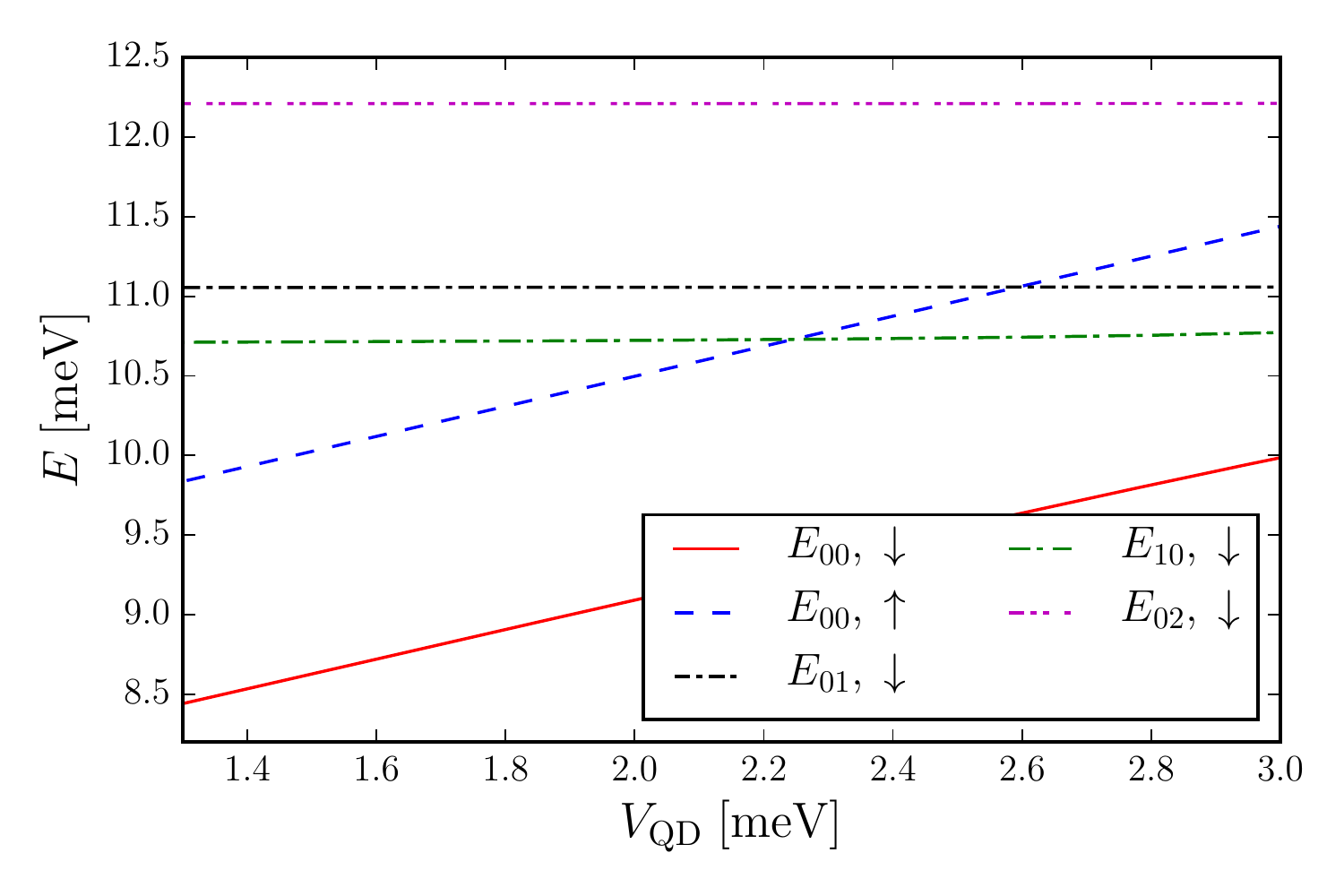}
\caption{Zeeman-split spin-up and spin-down energy levels of DRN as a function of $V_{\rm QD}$.}
\label{spc2}
\end{center}
\end{figure}

\begin{figure}[h]
\begin{center}
\includegraphics[width=0.7\textwidth]{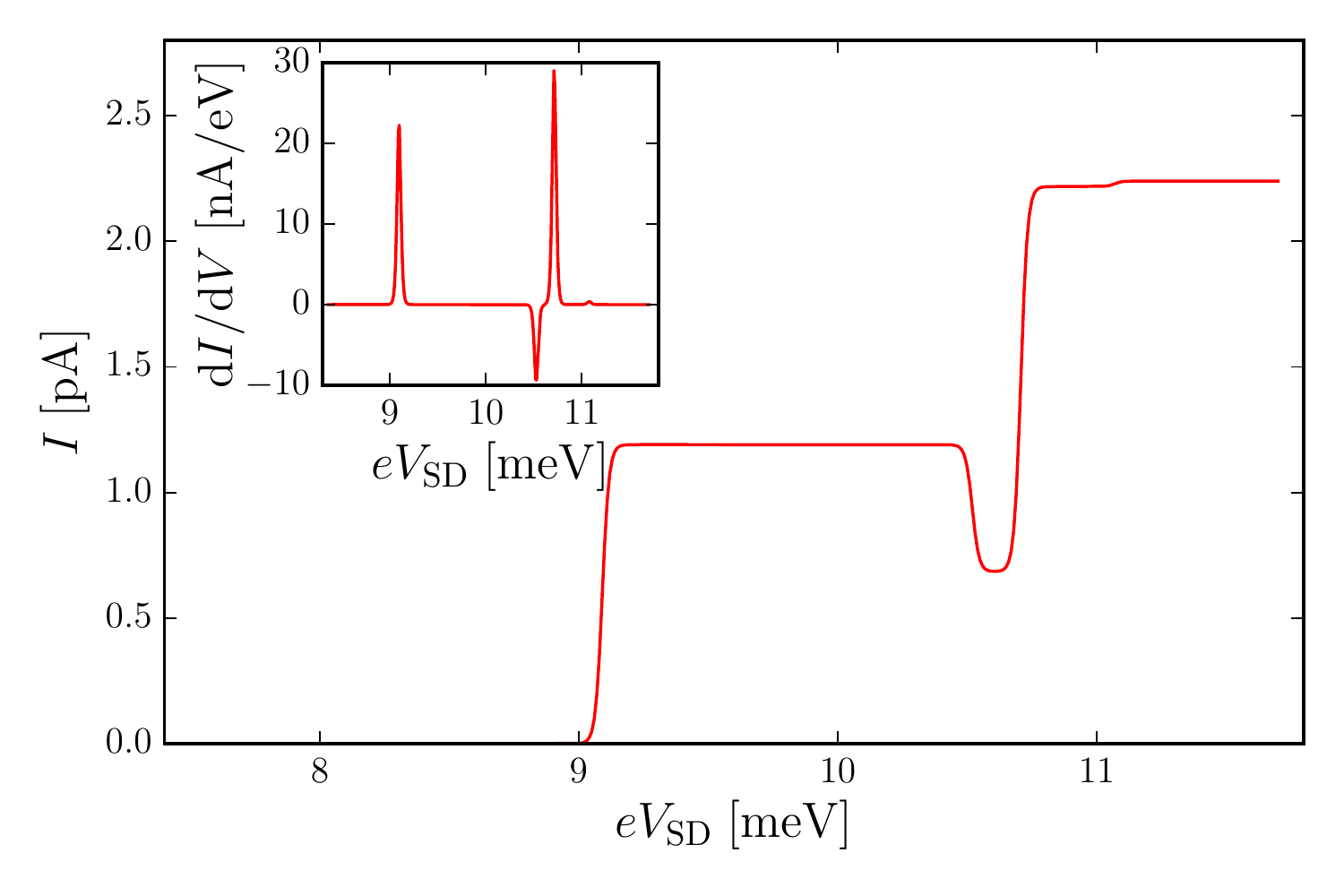}
\caption{Current as a function of applied source--drain voltage. The value of $V_{\rm QD}$ where NDC is pronounced is clearly visible in the inset, where the differential conductance ${\rm d}I/{\rm d} V$ is shown.
}
\label{c3}
\end{center}
\end{figure}

As a second example let us assume that the inner QD is made of material with $g_{\rm QD}=-4$ (e.g. InGaAs) and the outer QR is made of material with $g_{\rm QR}=-12$ (e.g. InAs). The energy spectrum as a function of $V_{\rm QD}$ is presented in Fig.~\ref{spc2}. and the dependence of the current (conductance) on the bias voltage $V_{\rm SD}$ with the pronounced NDC is shown in Fig.~\ref{c3}. 

\begin{figure}[h]
\begin{center}
\includegraphics[width=0.7\textwidth]{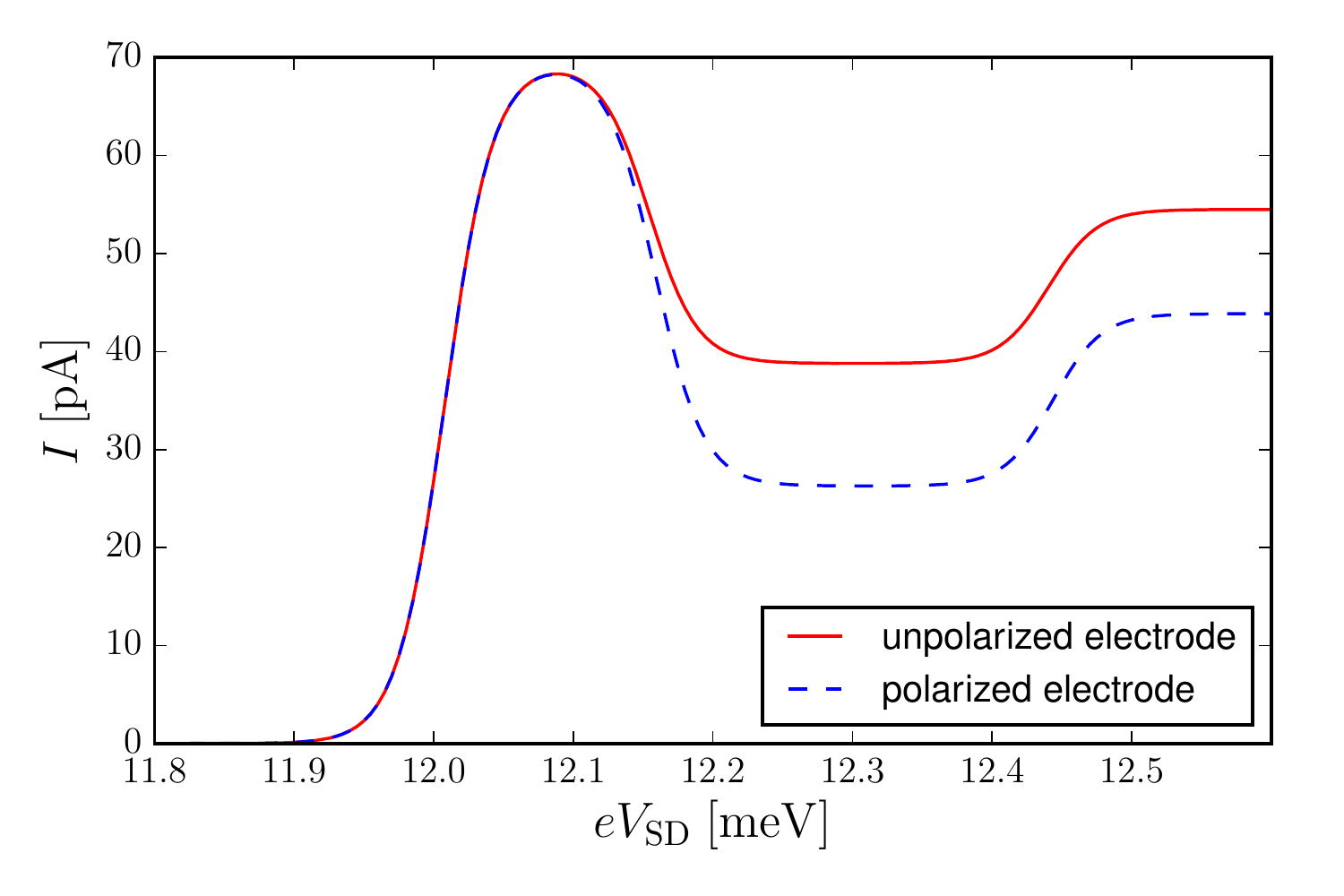}
\caption{Current as a function of applied source--drain voltage. Two curves refer to an unpolarized and polarized electrodes (solid red and dashed blue, respectively). NDC effect is enhanced for the case of spin polarized drain electrode.}
\label{drn_alfa}
\end{center}
\end{figure}

Finally, we consider the case illustrated in Fig.~\ref{lvl2} but with the drain electrode being spin polarized (Fig.~\ref{drn_alfa}). As expected, in this case the NDC effect is stronger due to the simultaneous influence of the unequal tunneling rates of $E_0^{\uparrow}$ and $E_0^{\downarrow}$ states and thicker
barrier for the spin-down electrons. It is shown in Fig.~\ref{drn_alfa} as a dashed blue curve. For comparison $I(V_{\rm SD})$ for unpolarised electrodes is also presented as a solid red line.

\section{Summary}\label{sec6}

Combined quantum structures are highly relevant to new technologies in which the control and manipulations of electron spin and wave functions play an important role.
 The ability to control the quantum state of a single electron is at the heart of many developments, e.g., in spintronics and quantum computing.
 
 In contrast to real atoms, DRNs allow flexible control over the confinement potential which gives rise to wave function engineering. We have performed systematic studies of such manipulations on conducting properties of dot-ring nanostructures. In particular, we have shown 
that it is possible to
 design a structure which exhibits spin related positive or negative differential resistance and switch between them by changing the gate voltage. The results indicate an opportunity to tune the performance of nanostructures and to optimize their specific properties by means of sophisticated structural design. Thus, the basic issues of quantum mechanics can be explored to design new semiconductor devices in which specific properties can be optimized.



\section*{Acknowledgements}
This work was supported by National Science Centre (NCN) under Grant No. DEC-2013/11/B/ST3/00824.

\bibliographystyle{unsrt}

\begin{thebibliography}{99}

\bibitem{hans} R. Hanson, L. P. Kouwenhoven, J. R. Petta, S. Tarucha, L. M. K. Vandersypen, Rev. Mod. Phys. {\bf 79}, 1217 (2007).

\bibitem{elzerman} J. M. Elzerman, R. Hanson, J. S. Greidanus, L. H. Willems van Beveren, S. De Franceschi, L. M. K. Vandersypen, S. Tarucha, L. P. Kouwenhoven Phys. Rev. B~{\bf 67}, 161308(R) (2003).

\bibitem{somaschini} C. Somaschini, S. Bietti, N. Koguchi and S. Sanguinetti, Nanotechnology {\bf 22}, 185602 (2011).

\bibitem{somaschini2} C. Somaschini and S. Bietti and N. Koguchi and S. Sanguinetti, Appl. Phys. Lett. {\bf 97}, 203109 (2010).

\bibitem{lauhon} L. J. Lauhon, M. S. Gudiksen, D. Wang, Ch. Lieber, Nature {\bf 420}, 57 (2002).

\bibitem{dillen} D. C. Dillen, K. Kim, E. Liu, E. Tutuc, Nature Nanotechnology {\bf 9}, 116 (2014).

\bibitem{mohseni} P. K. Mohseni, A. D. Rodrigues, J. C. Gaizerani, Y. A. Pusep, R. R. Pierre, Journ. of Appl. Phys. {\bf 106}, 124306 (2009).

\bibitem{sadowski} A. Siusys, J. Sadowski, M. Sawicki, S. Kret, T. Wojciechowski, K. Gas, W. Szuszkiewicz, A. Kami\'{n}ska, T. Story, Nanolett. {\bf 14}, 4263 (2014).

\bibitem{zipper} E. Zipper, M. Kurpas, and M. M. Ma\'{s}ka, New J. of Phys.14, 093029 (2012).

\bibitem{MK} M. Kurpas, E. Zipper, and M. M. Ma\'{s}ka, "Engineering of Electron States and Spin Relaxation in Quantum Rings and Quantum Dot-Ring Nanostructures" in "Physics of Quantum Rings", Vladimir M. Fomin (editor), Springer 2014, p. 455.

\bibitem{mk} M. Kurpas, B. K\k{e}dzierska, I. Janus-Zygmunt, A. Gorczyca-Goraj, E. Wach, E. Zipper, M. M. Ma\'{s}ka, J. Phys.: Condens. Matter {\bf 27}, 265801 (2015).

\bibitem{britnell} L. Britnell, R. V. Gorbachev, A. K. Geim,  L. A. Ponomarenko, A. Mishchenko, M. T. Greenway, T. M. Frombold, K. S. Novoselov, , L. Eaves, Nature Communications {\bf 4}, 1794 (2013).

\bibitem{bulka} T. Kostyrko, B. R. Bu{\l}ka, Phys. Rev. B {\bf 79}, 075310 (2009).

\bibitem{chen} J. Chen, M. A. Reed, A. M. Rawlett, J. M. Tour, Science {\bf 286}, 1550 (1999).

\bibitem{hawrylak} M. Ciorga, M. Pioro-Ladriere, P. Zawadzki, P. Hawrylak, A. S. Sachrajda, Appl. Phys. Lett. {\bf 80} 2177 (2002); M.Ciorga {\em et al}., Phys. Rev. B {\bf 61}, R16315 (2000).

\bibitem{biborski} A. Biborski, A. P. K\k{a}dzielawa, A. Gorczyca-Goraj, E. Zipper, M. M. Ma\'ska, and J. Spa\l{}ek, Sci. Rep. {\bf 6}, 29887 (2016).
 

\bibitem{Bardeen} J. Bardeen, Phys. Rev. Lett. {\bf 6}, 57 (1961).

\bibitem{bockelman} U. Bockelman and G. Bastard, Phys. Rev. B {\bf 42}, 8947 (1990).

\bibitem{nazarov} A. V. Khaetskii and Y. V. Nazarov, Phys. Rev B {\bf 64}, 125316 (2001).

\bibitem{Fujisawa} T. Fujisawa, Y.Tokura, Y.Hirayama, Phys. Rev. B~{\bf 63}, 081304(R), (2001).

\bibitem{wu} Y. Wu et al., ACS  Nano {\bf 6}, 2610 (2012).


\end{thebibliography}

\end{document}